\PassOptionsToPackage{english}{babel}
\documentclass[reprint,aps,pra,twocolumn,showpacs,amssymb,amsfonts,superscriptaddress,longbibliography]{revtex4-1}  

\usepackage[english]{babel}
\usepackage{graphicx}  
\usepackage{dcolumn}   
\usepackage{bm}        
\usepackage{bbm}        
\usepackage{amssymb}   
\usepackage{amsmath}
\usepackage{mathtools}
\usepackage[utf8]{inputenc}
\usepackage{xcolor}

\definecolor{myurlcolor}{rgb}{0,0,0.7}
\definecolor{myrefcolor}{rgb}{0.8,0,0}
\usepackage{hyperref}
\hypersetup{colorlinks, linkcolor=myrefcolor,
citecolor=myurlcolor, urlcolor=myurlcolor}

\newcommand{\avg}[1]{\left\langle #1\right\rangle}
\newcommand{\abs}[1]{\left\vert #1\right\vert}

\newcommand{\bra}[1]{\ensuremath{\langle#1|}}
\newcommand{\ket}[1]{\ensuremath{|#1\rangle}}

\newcommand{\ie}{\emph{i.e. }}
\newcommand{\eg}{\emph{e.g. }}

\begin{document}

\title{Number-phase entanglement and Einstein-Podolsky-Rosen steering}

\author{Matteo Fadel}
    \affiliation{State Key Laboratory for Mesoscopic Physics, School of Physics $\&$ Collaborative Innovation Center of Quantum Matter, Peking University, 100871 Beijing, China}
	\affiliation{Department of Physics, University of Basel, Klingelbergstrasse 82, 4056 Basel, Switzerland}

\author{Laura Ares}
    \affiliation{Departamento de \'{O}ptica, Facultad de Ciencias F\'{\i}sicas, Universidad Complutense, 28040 Madrid, Spain}

\author{Alfredo Luis}
    \affiliation{Departamento de \'{O}ptica, Facultad de Ciencias F\'{\i}sicas, Universidad Complutense, 28040 Madrid, Spain}

\author{Qiongyi He}\email{qiongyihe@pku.edu.cn}
	\affiliation{State Key Laboratory for Mesoscopic Physics, School of Physics $\&$ Collaborative Innovation Center of Quantum Matter, Peking University, 100871 Beijing, China}
	\affiliation{Beijing Academy of Quantum Information Sciences, Beijing 100193, China}
	\affiliation{Collaborative Innovation Center of Extreme Optics, Shanxi University, Taiyuan, Shanxi 030006, China}

\date{\today}

\begin{abstract}
We use the uncertainty relation between the operators associated to the total number of particles and to the relative phase of two bosonic modes to construct entanglement and Einstein-Podolsky-Rosen steering criteria. These can be tested experimentally in a variety of systems, such as optical fields, Bose-Einstein condensates or mechanical oscillators. While known entanglement criteria involving the phase observable typically require to perform interference measurements by recombining the two systems, our criteria can be tested through local measurements at two spatially distinct positions, to investigate the nonlocal nature of quantum correlations. We present simple examples where our criteria are violated, and show their robustness to noise. Apart from being useful for state characterization, they might find application in quantum information protocols, for example based on number-phase teleportation.
\end{abstract}

\maketitle

\section{Introduction} 
The Einstein-Podolsky-Rosen (EPR) paradox~\cite{EPR} occurs when measurements on one system allow to predict measurement results on an other system, with an accuracy that beats the limit posed by local uncertainty relations. The observation of such a paradox seems to imply that spatially separated measurements can influence each other irrespective of their separation, a mechanism that Schr\"odinger called ``steering"~\cite{Sch}.

From a conceptual point of view, the work by EPR reveals an inconsistency between our idea of local realism and the predictions of quantum mechanics. In the last decades, EPR steering has motivated numerous fundamental investigations as well as potential applications in quantum technologies~\cite{Reid89,EPRreview09}. 

Crucially, it has been shown that EPR steering is a distinctive manifestation of quantum correlations that differs from entanglement (state inseparability) \cite{Howard07}. In fact, EPR steering is a form of quantum nonlocality in which the roles of the involved parties are asymmetrical, and it enables the verification of shared entanglement even when one party's measurements are untrusted~\cite{Oneway12,Eric13,EPRreview17,EPRreview19}. This has a plethora of applications to one-sided-device-independent quantum communication~\cite{1sDIQKD,1sDIQKD_howard,HowardOptica,CV-QKDexp}, as well as to realize secure quantum teleportation~\cite{SQT13Reid,SQT15,SQT16_LiCM} and subchannel discrimination~\cite{subchannel,subchannel16,subchannelexp}.

A good number of experiments confirming the EPR paradox have been realized for mesoscopic optical fields by $XP$ quadrature measurements~\cite{EPRreview09,Ou92,Lee16,He19,ANU15,Su17,Cai20}. In the case of massive particles, entanglement between two spatially separated multipartite systems has been demonstrated for atomic ensembles at room temperature~\cite{Krauter2011,Julsgaard2011}, for Bose-Einstein condensates (BECs)~\cite{Fadel18,Kunkel18,Lange18}, and for mechanical oscillators~\cite{Riedinger18,Caspar18}. More recently, EPR steering was also observed in BECs~\cite{Peise15,Fadel18,Kunkel18}.

Criteria to detect entanglement and EPR steering strongly depends on the system (\eg continuous variable, spin), on the state preparation process, and on the measurement schemes that are available. Here we will focus on the case of bosonic modes (\eg optical or atomic), where states can be classified depending on the preparation processes into: i) non-number-conserving and ii) number-conserving. 

Examples for i) are the two-mode squeezed states originating from a pair-production process $H/\hbar= \kappa a^\dagger b^\dagger +\kappa^* ab$, where $\langle ab\rangle\neq 0$ but $\langle a^\dagger b\rangle=0$. These are typical states prepared in optics via parametric down-conversion~\cite{Reid89,EPRreview09} or nondegenerate four-wave mixing~\cite{fwm}, and in BECs via spin exchanging collisions \cite{Peise15}. 

On the other hand, examples for ii) are states originating from a beam-splitter operation $H/\hbar= \kappa a^\dagger b+\kappa^* ab^\dagger$, so that $\langle ab\rangle= 0$ but $\langle a^\dagger b\rangle\neq0$. These are typical states prepared in optics via linear beam-splitters, and in double-well BECs through tunneling dynamics \cite{Esteve08}. 

Entanglement and EPR steering can be detected in case i) through criteria based on local measurements of the harmonic oscillator $XP$ quadratures. For mode $a$ these are defined as $X_A=(a^\dagger+a)/\sqrt{2}$ and $Y_A=(a^\dagger -a)/i\sqrt{2}$, and a similar definition holds for mode $b$. These quadratures are measured experimentally through homodyne detection, where each mode is interfered with a local oscillator that serves as a phase reference. Remarkably, apart from the optical case~\cite{Ou92}, this has also been demonstrated in atomic~\cite{Peise15} systems.

On the other hand, criteria for entanglement and EPR steering based on $XP$ quadrature measurements are not suited to states in case ii), due to $\langle ab\rangle = 0$. Nevertheless, one can use other criteria, such as the Hillery-Zubairy (HZ) non-Hermitian operator product criterion~\cite{HZ} stating that a violation of $|\langle a^\dagger b\rangle |^2 \leq \langle a^\dagger a  b^\dagger b\rangle$ implies that the modes $a$ and $b$ are entangled. A generalization of such inequality can also be used to formulate HZ-type criteria for EPR steering~\cite{He2012,Eric2011}, confirming that mode $a$ is steered by mode $b$ if $|\langle a^\dagger b\rangle |^2>\langle a^\dagger a  (b^\dagger b+1/2)\rangle$, or mode $b$ is steered by mode $a$ if $|\langle a^\dagger b\rangle |^2>\langle (a^\dagger a+1/2) b^\dagger b\rangle$.

However, note that the types of criteria just mentioned require measurements that do not have a clear interpretation in terms of local observables that could be addressed at spatially separated positions \cite{He2012,Geza03,Inigo10}. In fact, terms like $\langle a^\dagger b\rangle$ consists in interference measurements that require to recombine the two modes, and are therefore nonlocal measurements. While in many practical situations such measurements can be legitimate for state characterization, in general they cannot be used to rigorously investigate the nonlocal nature of quantum correlations, or for state/device-independent quantum information tasks.

In this paper we present new criteria to detect entanglement and EPR steering between two spatially separated bosonic modes, that are based on local measurement of the conjugate number and phase observables.
Since the definition of a phase operator in quantum mechanics is notoriously non trivial \cite{Nieto,Lynch95}, we pay particular attention to address this complication rigorously. In fact, as there is actually no such well defined operator \cite{Nieto,phase,phase2}, we follow the most general quantum description of an observable in terms of a Positive-Operator Valued Measure (POVM). Moreover, we quantify the phase uncertainty in terms of the so called dispersion which, contrary to the variance, is tailored to angular variables.
Having these tools defined, we then derive criteria to test entanglement and EPR steering based on the number-phase uncertainty relation \cite{TO,TO2}. More specifically, we consider the number sum and phase difference as the two basic compatible observables whose uncertainty is bounded from below for separable or non-steerable states.

\section{Intuitive approach} 

Consider two systems, labeled by $j=1,2$, on which measurements $A_j$ and $B_j$, with $[A_{j},B_{j\prime}]=\delta_{j,j\prime} C_{j}$, are performed. For all separable states between the two systems it is known to hold the relation~\cite{DGCZ,Simon}
\begin{equation}\label{generalENT}
    \Delta^2(A_1+A_2) + \Delta^2(B_1-B_2) \geq \left( |\langle C_1\rangle| + |\langle C_2\rangle| \right) \;.
\end{equation}
Here, $\Delta^2(X)=\langle X^2\rangle-\langle X\rangle^2$ is the variance of the operator $X$. Similarly, for non-steerable states it holds~\cite{Reid89,EPRreview09}
\begin{equation}\label{generalEPR}
    \Delta^2(A_1+A_2) + \Delta^2(B_1-B_2) \geq |\langle C_2\rangle| \;.
\end{equation}
A typical choice of measurements is position and momentum operators, $A_j=X_j$, $B_j=P_j$, for which $C_j=i$. This has allowed to detect entanglement and steering in continuous variable systems~\cite{EPRreview09,He19,Ou92,Oneway12,ANU15,Su17,Cai20}. An other possibility is to chose spin observables, whose commutator is now also an operator~\cite{Bowen,VG}. This has allowed to detect entanglement and steering  between atomic ensembles~\cite{Krauter2011,Julsgaard2011,Fadel18,Kunkel18,Lange18}.

When considering bosonic modes, one can also be tempted to choose for $A_j$ and $B_j$ the particle number operators $N_j$ and their conjugate phase operators $\phi_j$ (we will discuss the subtleties of this latter in the following paragraph). Naively, these number and phase operators are expected to satisfy the canonical commutation relation~\cite{Dirac,Nieto}
\begin{equation}\label{naiveNPHUR}
    \left[ N_j, \phi_j \right] = i \qquad\text{(in general wrong~\cite{Nieto})} \;,
\end{equation}
and therefore to satisfy the uncertainty relation
\begin{equation}\label{naiveNPprod}
    \Delta^2 N_j \Delta^2 \phi_j \geq \dfrac{1}{4} \qquad\text{(in general wrong)} \;.
\end{equation}
or alternatively, as $x^2 + y^2 \geq 2 \sqrt{ x^2 y^2 }$, also
\begin{equation}\label{naiveNPsum}
    \Delta^2 N_j + \Delta^2 \phi_j \geq 1 \qquad\text{(in general wrong)} \;.
\end{equation}
From these relations and Eqs.~(\ref{generalENT},\ref{generalEPR}), we are expected to certify entanglement if it is violated the inequality
\begin{equation}\label{naiveENT}
    \Delta^2(N_1+N_2) + \Delta^2(\phi_1-\phi_2) \geq 2 \;,
\end{equation}
and steering if it is violated the inequality
\begin{equation}\label{naiveEPR}
    \Delta^2(N_1+N_2) + \Delta^2(\phi_1-\phi_2) \geq 1 \;.
\end{equation}
Note that these criteria involve the total-number operator $N:=N_1+N_2$ and the phase-difference operator $\phi:=\phi_1-\phi_2$. These would allow to detect correlations in observables that are not the usual $XP$ quadratures, therefore characterizing a different class of states.

Unfortunately, as it was mentioned before, the definition of a phase operator in quantum mechanics is a subtle task, which makes the expressions presented so far to be in general wrong~\cite{Nieto,Lynch95}. In fact, while approximate operators satisfying Eqs.~(\ref{naiveNPHUR}) and (\ref{naiveNPprod}) can be found in the limit of small phase fluctuations, this is not true in a more general case.

In the following we will treat this problem rigorously, to derive entanglement and steering criteria that are valid for arbitrarily large phase fluctuations and that are experimentally practical. As expected, in the limit of small phase fluctuations our criteria allow to recover Eqs.~(\ref{naiveENT}) and (\ref{naiveEPR}) from a rigorous framework.

\bigskip
\section{Number-Phase observables} 

In this section we introduce the operators associated to the total number of particles and to the relative phase of two bosonic modes. We discuss their properties, their eigenstates, and how to express their fluctuations. To conclude, we present the uncertainty relation holding in these observables, which will later be of central importance for deriving entanglement and steering criteria.

\subsection{Number and Phase operators} 

We are interested in investigating correlations between number and phase observables. In classical physics these two observables arise naturally in the context of \eg oscillating fields. In quantum mechanics, however, the definition of a phase operator is less straightforward \cite{phase,phase2}.

For a single bosonic mode defined by the operator $a_j$, the total number of particles is simply $N_j:=a_j^\dagger a_j$. A physically meaningful choice is to describe single-mode phase via the Positive-Operator Valued Measure (POVM)
\begin{equation}\label{DpPOVM}
    \Pi_j (\phi) = \ket{\phi}_j \bra{\phi}_j \;,
\end{equation}
with the non-normalizable, non-orthogonal phase states
\begin{equation}
\label{phst}
    \ket{\phi}_j = \frac{1}{\sqrt{2\pi}}\sum_{n=0}^\infty e^{i n \phi} \ket{n}_j \;.
\end{equation}
These latter are unit-modulus-eigenvalue eigenstates of the Susskind-Glogower \cite{SuskGlog} exponential-of-phase operator $E_j$, namely $E_j \ket{\phi}_j= e^{i \phi} \ket{\phi}_j$, such that
\begin{equation}\label{SGEopInt}
E_j = \int_{2\pi} \text{d} \phi \; e^{i\phi} \Pi_j (\phi) \;.
\end{equation}
For later reference, note that this operator can also be seen as a ``normalized'' ladder operator, namely $E_j= \sum_{n=0}^\infty \ket{n}\bra{n+1}=(N_j+1)^{-1/2} a_j$.

In the following we will be interested in a system constituted by two bosonic modes, defined by the operators $a_1$ and $a_2$. Inspired by the single mode case we first define a total number operator as $N:= N_1 + N_2$. Then, we introduce an operator associated to the relative phase between the two modes, say $\phi := \phi_1 - \phi_2$. To this end, let us first construct the joint POVM for the two single-mode phases $\phi_1$ and $\phi_2$ as 
\begin{equation}
    \Pi (\phi_1 , \phi_2 ) = \Pi_1 (\phi_1) \otimes \Pi_2 (\phi_2) \;.  
\end{equation}
Since here we are only interested in the relative phase, we may consider the change of variables 
\begin{equation}
   \phi = \phi_1 - \phi_2 \;, \qquad \varphi= \phi_2 \;,
\end{equation}
so that
\begin{equation}
    \Pi (\phi , \varphi ) = \Pi_1 (\phi+\varphi) \otimes \Pi_2 (\varphi) \;.
\end{equation}
From this expression, we finally obtain the POVM associated to the phase difference $\Pi(\phi)$ by integrating out the variable $\varphi$ as
\begin{equation}\label{Dp}
    \Pi(\phi) =\int_{2\pi} \text{d}\varphi\;  \Pi (\phi , \varphi ) = \int_{2\pi} \text{d} \varphi\; \Pi_1 (\phi+\varphi) \otimes \Pi_2 (\varphi) \;.  
\end{equation}
Inserting in this expression Eqs.~(\ref{DpPOVM},\ref{phst}) for the single-mode phase states, and performing the integration over $\varphi$, we arrive at 
\color{black}
\begin{equation}\label{DpDiffPOVM}
   \Pi (\phi) = \dfrac{N+1}{2 \pi} \sum_{N=0}^\infty \ket{N, \phi}\bra{N, \phi} \;,
\end{equation}
where $\ket{N, \phi}$ are now the normalized, non-orthogonal number-phase states
\begin{equation}\label{nphi}
\ket{N, \phi} = \frac{1}{\sqrt{N+1}} \sum_{m=0}^N e^{i m \phi} \ket{m}_1 \ket{N-m}_2 \;.
\end{equation}
These latter are unit-modulus-eigenvalue eigenstates of the exponential-of-phase-difference operator $E$, namely $E \ket{N, \phi} = e^{i\phi} \ket{N, \phi}$, such that
\begin{equation}\label{EE1E2}
E = \int_{2\pi} \text{d}\phi\; \;e^{i\phi} \Pi (\phi) = E_1 E^\dagger_2 \;.
\end{equation}
Moreover, the states $\ket{N, \phi}$ are also eigenstates of the total number operator with eigenvalue $N$. This observation reflects the expected compatibility between total number and phase difference, which means $[N, \Pi (\phi)] = 0$.
Similarly to $E_j$, it is interesting to mention that $E$ can be related to the ladder operators $a_j$ as $E=((N_1+1)N_2)^{-1/2}a_1 a_2^\dagger$ \cite{phase}. This relation illustrates how $E$ depends on the coherence between the two bosonic modes.

It is important to emphasize that the relation Eq.~\eqref{Dp} expresses the idea that the relative phase $\phi$ between the two modes can be determined via independent local measurements of $\phi_1$ and $\phi_2$ on the respective mode. This is because Eq.~\eqref{Dp} implies that the probability distribution $p(\phi)$ for measuring $\phi$ is obtained from the joint probability distribution $p(\phi_1,\phi_2)$ by summing over all configurations for which $\phi_1-\phi_2=\phi$, namely we have that
\begin{equation}
 p (\phi ) = \int_{2\pi} \text{d}\varphi\; p(\phi_1 =\phi+\varphi,\phi_2 = \varphi) \;.  
\end{equation}
Similarly, Eq.~\eqref{EE1E2} reflects also this fact by showing that the relative phase operator $E$ can be expressed from the single mode phase operators $E_1$ and $E_2$. Concretely, the local POVMs associated to the single mode phases, $\Pi_j (\phi_j)$, can be implemented experimentally as projective measurements following the prescription given by the Naimark extension \cite{Naimark43,Pregnell02,Pozza19}.

In the following, for the sake of readability, we will often call the total number and phase difference operators simply number and phase operators.

\subsection{Number Variance and Phase Dispersion}

The formulation of an uncertainty relation between number and phase operators requires a quantification of their fluctuations. For the number observable this is simply achieved by considering the variance 
\begin{equation}\label{varN}
    \Delta^2 N = \avg{N^2} - \avg{N}^2 \;.
\end{equation}
However, fluctuations in the phase are not properly characterized by the ``standard'' definition of variance as in Eq.~\eqref{varN}. What it is done instead, is to define the so called called phase dispersion $D^2$ \cite{TO,TO2,Breitenberger89}. For a single mode this is computed as
\begin{equation}\label{Dj2def}
    D^2_j = 1 - \abs{\avg{ E_j }}^2 \;, 
\end{equation}
while for two modes the disperion of the phase difference is (remember $E=E_1 E_2^\dagger$)
\begin{equation}\label{D2def}
     D^2 = 1 - \abs{\avg{ E }}^2 \;.
\end{equation}
While the variance in Eq.~\eqref{varN} is only bounded to be non-negative, for the dispersion it holds $1 \geq D_j, D \geq 0$, where zero corresponds to no phase fluctuations and unity to uniform phase distribution. To emphasize the fact that Eq.~\eqref{Dj2def} is associated to the phase of a single mode, while Eq.~\eqref{D2def} is associated to the phase difference between two modes, in the following we will call $D_j^2$ the phase dispersion and $D^2$ the relative-phase dispersion.

In the limit of small phase fluctuations, the probability distribution for the phase $P(\phi)$ will be peaked around some mean value that, without loss of generality, we can consider to be zero. Under this limit, a series expansion of the exponential-of-phase operators is valid. Starting from the relation $D^2 = 1 - \abs{\avg{e^{i\phi}}}^2$ it is immediate to see that (to second order)
\begin{equation}
\label{spf}
    D^2 \simeq \langle\phi^2\rangle - \langle\phi\rangle^2 \equiv \Delta^2 \phi \;.
\end{equation}
Here $\Delta^2 \phi$ has the usual meaning of variance for the probability distribution for the phase $P(\phi)$.

\subsection{Number-Phase Uncertainty Relations}

\begin{figure}
  \centering
\includegraphics[width=0.45\textwidth]{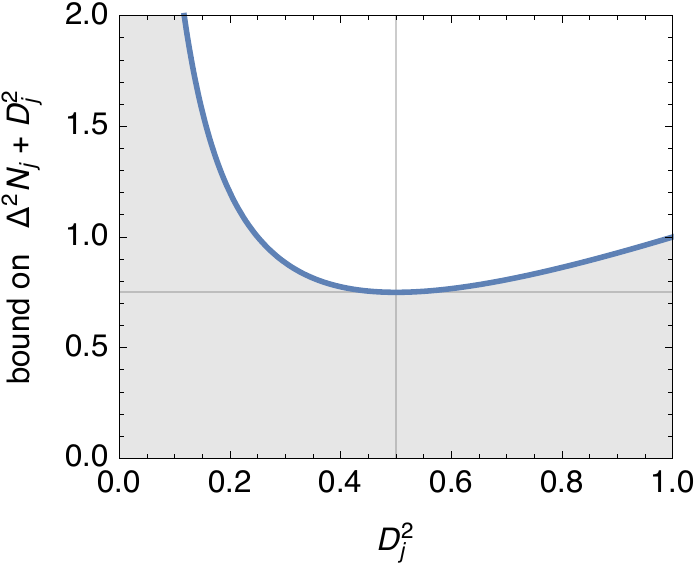}
  \caption{Evaluation of the bound for $\Delta^2 N_j + D_j^2$, Eq.~\eqref{smur}. The gray region below the blue curve is forbidden by the uncertainty relation. The horizontal gray line at $3/4$ indicates the state-independent bound which is tight for $D_j^2 = 1/2$ (vertical gray line). 
  }\label{figNDsumBound}
\end{figure}

Derivations of entanglement and EPR-steering criteria are often based on the uncertainty relations between the considered observables. 
For number-phase observables in a single mode we consider the uncertainty relation presented in Ref.~\cite{TO}, which reads
\begin{equation}\label{ND}
\left ( \Delta^2 N_j + \frac{1}{4} \right ) D_j^2 \geq \frac{1}{4} \;.
\end{equation}
This relation can also be written as
\begin{equation}\label{smur}
\Delta^2 N_j +  D_j^2 \geq \dfrac{1}{4 D_j^2} - \dfrac{1}{4} + D_j^2 \geq \frac{3}{4} \;,
\end{equation}
where the constant $3/4$ has been found by minimizing the term in the middle, see Fig.~\ref{figNDsumBound}.
Note here that this constant lower bound is in general not tight, as it is attained only for a state with $\Delta^2 N_j = 1/4$ and $D_j^2 = 1/2$.

In the case of small phase fluctuations we might expect to recover the uncertainty relations Eqs.~(\ref{naiveNPprod},\ref{naiveNPsum}). To see that this is the case, we start by rewriting Eq.~\eqref{ND} as
\begin{equation}
\label{URprod}
\Delta^2 N_j \frac{D_j^2}{1 - D_j^2} \geq \dfrac{1}{4} \;.
\end{equation}
Because for small phase fluctuations $D_j^2\approx 0$, it holds the series expansion
\begin{equation}\label{ratioDsmall}
    \dfrac{D_j^2}{ 1 - D_j^2 } \simeq D_j^2 \simeq \Delta^2 \phi_j \;,
\end{equation}
such that the ``naive'' uncertainty relation Eq.~\eqref{naiveNPprod} is recovered. From this, and the triangle inequality, we immediately recover also Eq.~\eqref{naiveNPsum}. 

To conclude let us remember that, in the case of two modes, total number and phase difference are compatible observables, resulting in a trivial uncertainty relation.


\section{Entanglement criterion}

Our goal here is to derive an entanglement criterion based on total-number and phase-difference observables. As these associated operators commute, for all quantum states it holds the trivial inequality
\begin{equation}
\Delta^2 N +  D^2 \geq 0 \;.     
\end{equation}
However, if we restrict ourselves to separable states of the two modes, we are able to provide a non-zero lower bound for $\Delta^2 N +  D^2$. The idea behind our proof follows the approach used in Ref.~\cite{DGCZ,Simon,Hofmann}.

In full generality, separable bipartite states can be written as the convex combination
\begin{equation}\label{sep}
    \rho_\mathrm{sep} = \sum_k p_k \; \rho_{1,k} \otimes \rho_{2,k} \;,
\end{equation}
where $p_k \geq 0$, $\sum_k p_k = 1$, and $\rho_{j,k}$ is a density matrix for mode $j$. When evaluated on the separable state $\rho_{1,k} \otimes \rho_{2,k}$, the number variance is \begin{equation}\label{Ncocn2}
    \Delta^2_k N = \Delta^2_k N_{1} + \Delta^2_k N_{2} \;.
\end{equation}
For the phase dispersion the decomposition is more subtle. Separability implies $| \langle E_1 E^\dagger_2 \rangle_k |^2 = | \langle E_1 \rangle_k |^2 | \langle E_2^\dagger \rangle_k |^2 = | \langle E_1 \rangle_k |^2 | \langle E_2 \rangle_k |^2$, and we obtain
\begin{equation}\label{Dcocn1}
    D_k^2 = D_{k,1}^2 + D_{k,2}^2 - D_{k,1}^2 D_{k,2}^2 \;. 
\end{equation}

Using Eqs.~(\ref{Ncocn2},\ref{Dcocn1}), we find that for separable states
\begin{align}
& \left( \Delta_k^2 N + 1 \right) D_k^2  = \nonumber\\
&\; = \left( \Delta_k^2 N_1 + \frac{1}{4} \right) D_k^2 + \left( \Delta_k^2 N_2 + \frac{1}{4} \right) D_k^2 + \frac{D_k^2}{2} \nonumber\\
&\; \geq \frac{1}{2} + \left( \Delta_k^2 N_1 + \frac{1}{4} \right) D_{k,2}^2 \left(1-D_{k,1}^2\right) + \nonumber\\
&\; \phantom{asd} + \left( \Delta_k^2 N_2 + \frac{1}{4} \right) D_{k,1}^2 \left(1-D_{k,2}^2\right) + \frac{D_k^2}{2} \nonumber\\
&\; \geq \frac{1}{2} + \dfrac{D_{k,2}^2 \left(1-D_{k,1}^2\right)}{4 D_{k,1}^2} +  \dfrac{D_{k,1}^2  \left(1-D_{k,2}^2\right)}{4 D_{k,2}^2} + \frac{D_k^2}{2} \nonumber\\
&=\frac{1}{2}+\frac{\left(D_{k,2}^{2}\right)^{2}+\left(D_{k,1}^{2}\right)^{2}}{4 D_{k,1}^{2} D_{k,2}^{2}}+\frac{D_{k,1}^{2}+D_{k,2}^{2}-2 D_{k,1}^{2} D_{k,2}^{2}}{4} \nonumber\\
&\geq\frac{1}{2}+\frac{2}{4} = 1 \;, \label{ENTkBound}
\end{align}
where to derive the first two inequalities we used the uncertainty relation for each system, Eq.~\eqref{ND}, and in going to the last line we used the triangle inequality  
\begin{equation}
\left(D_{k,1}^{2}\right)^{2}+\left(D_{k,2}^{2}\right)^{2} \geq 2 D_{k,1}^{2} D_{k,2}^{2} \;,
\end{equation} 
and
\begin{equation}
D_{k,1}^{2}+D_{k,2}^{2} \geq 2 D_{k,1} D_{k,2} \geq 2 D_{k,1}^2 D_{k,2}^{2} \;.
\end{equation}

Since for all states Eq.~\eqref{sep} we have for the variance $\Delta^2 N \geq \sum_k p_k \Delta^2_k N$, and for the relative-phase dispersion $D^2 \geq \sum_k p_k D^2_k$, we can prove that for all separable states
\begin{align}
 \left ( \Delta^2 N + 1 \right ) D^2  &\geq \left( \sum_k p_k (\Delta_k^2 N + 1 ) \right) \sum_k p_k D_k^2 \nonumber\\
&\geq \left[ \sum_k p_k \sqrt{\left( \Delta_k^2 N + 1 \right) D_k^2} \right]^2 \nonumber\\
&  \geq \left[ \sum_k p_k \sqrt{1} \right]^2 = 1 \;. \label{NDentProof}
\end{align}
Here, in going to the third line we used the Cauchy-Schwarz inequality, and in going to the fourth we used Eq.~\eqref{ENTkBound}.

\begin{figure}
  \centering
\includegraphics[width=0.45\textwidth]{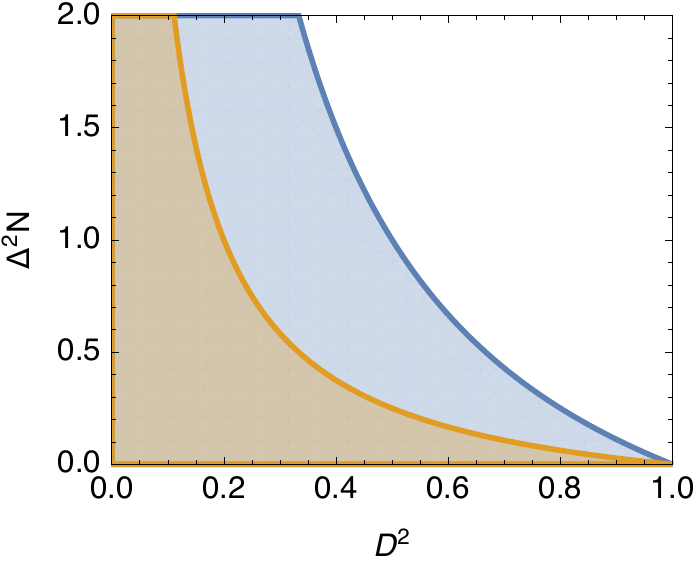}
  \caption{States in the parameter region below the blue (upper) curve violate inequality Eq.~(\ref{NPent}), meaning that they are entangled. States in the parameter region below the orange (lower) curve violate (\ref{NPepr}), meaning that they are two-way steerable. }\label{figCRIT}
\end{figure}

To summarize, as we proved that for all separable states it holds
\begin{equation}\label{NPent}
\left ( \Delta^2 N + 1 \right ) D^2 \geq 1 \;,
\end{equation}
any violation of this inequality certifies entanglement between the two modes. Therefore, Eq.~\eqref{NPent} is a bipartite entanglement criterion involving the total number of particles $N$ and the relative-phase dispersion $D^2$ associated to the phase difference between the two modes.
For illustration purposes, Fig.~\ref{figCRIT} shows in blue the parameter region for which Eq.~\eqref{NPent} is violated. States associated to this region are therefore entangled.

In the limit of small fluctuations in the phase difference, where an expression analogous to Eq.~\eqref{ratioDsmall} holds for $D^2$, we obtain from Eq.~\eqref{NPent} the entanglement criterion
\begin{equation}
\Delta^2 N \frac{D^2}{1 - D^2} \approx \Delta^2 N \Delta^2 \phi \geq 1 \;.
\end{equation}
From this, the triangle inequality implies $\Delta^2 N + \Delta^2 \phi \geq 2 \sqrt{\Delta^2 N \Delta^2 \phi} \geq 2$, which is the entanglement criterion we were expecting for the sum of variances Eq.~\eqref{naiveENT}.
%


\section{Einstein-Podolsky-Rosen steering criterion}

The general idea behind a derivation of an EPR steering criterion follows the same approach as the one for entanglement, but we use the quantum uncertainty relation only for the system that has been assumed to be a ``local quantum state"~\cite{Howard07}. For the other system (\ie for the system that steers), we do not assume anything about the variances of the local states except that they are positive. 

Inspired by Eq.~\eqref{ENTkBound} we start from a similar expression and, using again Eqs.~(\ref{Ncocn2}) and (\ref{Dcocn1}), we find that for all non-steerable states
\begin{align}
& \left( \Delta_k^2 N + \frac{1}{4} \right) D_k^2  = \nonumber\\
& \phantom{asd} = \left( \Delta_k^2 N_1 + \Delta_k^2 N_2 + \frac{1}{4} \right) D_k^2 \nonumber\\
& \phantom{asd} = \Delta_k^2 N_1 D_k^2 + \left( \Delta_k^2 N_2 + \frac{1}{4} \right) \left( D_{k,1}^2 + D_{k,2}^2 - D_{k,1}^2D_{k,2}^2 \right) \nonumber\\
& \phantom{asd} \geq \Delta_k^2 N_1 D_k^2 + \left( \Delta_k^2 N_2 + \frac{1}{4} \right)  D_{k,1}^2 \left( 1 - D_{k,2}^2 \right) + \frac{1}{4} \nonumber\\
& \phantom{asd} \geq \frac{1}{4} \;,
\end{align}
where in going to the second to last line we used the uncertainty relation for system $2$ (steered party), and in going to the last line we used the fact that for system $1$ no uncertainty relation applies, meaning that we can set simultaneously $\Delta_k^2 N_1=D_{k,1}^2=0$. Note here that the same result could have been obtained also for the opposite choice in the uncertainty bounds, correspondig to the situation where system 1 is steered.

Following the same steps as in Eq.~\eqref{NDentProof}, we obtain that for all non-steerable states it holds
\begin{equation}\label{NPepr}
\left ( \Delta^2 N + \frac{1}{4} \right ) D^2 \geq \frac{1}{4} \;.
\end{equation}
Therefore, this inequality is a bipartite steering criterion involving the total number of particles $N$ and the relative-phase dispersion $D^2$, whose violation actually implies two-way steering between the systems.

For illustration purposes, Fig.~\ref{figCRIT} shows in orange the parameter region for which Eq.~\eqref{NPepr} is violated. States associated to this region are therefore (two-way) steerable. Moreover, note that Fig.~\ref{figCRIT} highlights the hierarchy existing between entanglement and steering: steering is a stronger form of correlation, for which entanglement is necessary but not sufficient. As a consequence, every state showing steering is necessarily entangled.

In the limit of small fluctuations in the phase difference, where an expression analogous to Eq.~\eqref{ratioDsmall} holds for $D^2$, we obtain from Eq.~\eqref{NPepr} the steering criterion
\begin{equation}
\Delta^2 N \frac{D^2}{1 - D^2} \approx \Delta^2 N \Delta^2 \phi \geq \dfrac{1}{4} \;.
\end{equation}
From this, the triangle inequality implies $\Delta^2 N + \Delta^2 \phi \geq 2 \sqrt{\Delta^2 N \Delta^2 \phi} \geq 1$, which is the steering criterion we were expecting for the sum of variances Eq.~\eqref{naiveEPR}.

\section{Examples}

In this section we analyse a number of experimentally relevant examples, to illustrate the usefulness of the number-phase entanglement and steering criteria we derived.

\subsection{Number-phase states} 
As a first example we consider the number-phase states described in Eq.~\eqref{nphi}, for which we have
\begin{equation}
\Delta^2 N = 0 \;,\qquad  D^2 = \frac{2N+1}{(N+1)^2} \;.
\end{equation}
Here, the second equality comes from the fact that $\avg{E}=N/(N+1)$.
As the total number of particles is constant, both entanglement and EPR criteria, Eqs.~(\ref{NPent},\ref{NPepr}), reduce to  $D^2 \geq 1$. However, since $0\leq D^2\leq 1$, the latter coincides with the condition
\begin{equation}
    D^2 = 1 \;,
\end{equation}
which is violated whenever $N>0$.

From the previous observations, we conclude that when the number of particles is fixed, $\Delta^2 N =0$ our entanglement and EPR criteria always reduce to $D^2 = 1$, see Fig.~\ref{figCRIT}. Therefore, every state violating this condition shows directly both entanglement and steering.

\subsection{Split Fock states}
States showing mode entanglement can be prepared using beam splitters. For example, let consider the input state to be the product state of a Fock state and vacuum, \ie $\ket{N}\ket{0}$. Then, the output of the beam splitter are $SU(2)$ coherent states. For a balanced beam splitter where the splitting ratio is $50:50$, the output state in the number basis reads
\begin{equation}\label{cps}
    |N,\varphi \rangle =\frac{1}{\sqrt{2^N}} \sum_{m=0}^N \sqrt{\left ( \begin{array}{c} N  \\ m \end{array} \right )} e^{im \varphi}|m\rangle|N-m \rangle \;,
\end{equation}
where $\varphi$ is some relative phase introduced by the beam splitter.

These states have $\Delta^2 N =0$, as the number of particles is set by the input Fock state, while the relative-phase dispersion is
\begin{equation}
D^2=1-\left[\dfrac{1}{2^N} \sum^N_{m=1}\sqrt{\left ( \begin{array}{c} N  \\ m \end{array} \right )\left ( \begin{array}{c} N  \\ m-1 \end{array} \right ) }\right]^2 \;.
\end{equation}
Like in the previous example, as there are no $N$ fluctuations, entanglement and EPR criteria reduces to $D^2 = 1$, which is violated whenever $N>0$.

For this example it is straightforward to compare our criteria with the HZ-type criteria mentioned in the introduction \cite{HZ,He2012,Eric2011}. A simple calculation yields $\langle a^\dagger a \rangle = \langle b^\dagger b \rangle = N/2$, $\left | \langle a^\dagger b \rangle \right |^2 = N^2/4$, and $\langle a^\dagger a b^\dagger b \rangle = N(N-1)/4$. These show that the separability criterion $|\langle a^\dagger b\rangle |^2 \leq \langle a^\dagger a  b^\dagger b\rangle$ is violated for all $N$, detecting entanglement. On the contrary, the criteria for non-steerable states, \eg $|\langle a^\dagger b\rangle |^2 \leq \langle a^\dagger a  (b^\dagger b+1/2)\rangle$, are never violated. Therefore, this example emphasizes the fact that the number-phase criteria we derived allow us to detect steering in classes of states in which other HZ-type criteria cannot.

\subsection{Two-mode squeezed states}
Nonclassical states that are known to show entanglement and steering are two-mode squeezed states (TMSS). These are prepared in optical experiments using parametric down-conversion. In the number basis, TMSS read
\begin{equation}\label{tmss}
    |\psi(r) \rangle = \frac{1}{\cosh{r}} \sum_{m=0}^\infty (\tanh{r})^m |m\rangle|m \rangle \;,
\end{equation}
where $r\geq 0$ is a real parameter associated with the squeezing strength. For these states we obtain
\begin{equation}
\Delta^2 N = \sinh^2(2r) \;,\qquad  D^2 =1  \;,
\end{equation}
implying that both our entanglement and steering criteria, Eqs.~(\ref{NPent}) and (\ref{NPepr}), are never violated. 

Despite this result, let us remember that TMSS show a violation of entanglement an steering criteria based on $XP$ quadrature measurements~\cite{Reid89,EPRreview09} which, on the other hand, are useless for number-phase states or split Fock states. Therefore, this example emphasizes the fact that the number-phase criteria we derived  allow us to detect entanglement and steering in classes of states in which standard $XP$ criteria cannot.

\subsection{States with noise in the total number}

So far we have discussed pure states, which represent idealized situations. In fact, experiments always deal with different types of noise. We analyze here the effect of an imperfect state preparation resulting in shot-to-shot fluctuations in the total number of particles.

The expectation value of an operator $A$ is
\begin{equation}
\langle A \rangle = \sum_{N=0}^\infty \mathrm{tr} \left ( \rho_N A \right ) \;,
\end{equation}
where $\rho_N$ is the restriction of $\rho$ to the subspace with total number $N$.

To give a concrete example, consider a statistical mixture of number-phase states Eq.~\eqref{nphi}. (Same results have been obtained numerically for the split Fock states (\ref{cps})).
We have $\rho_N = p(N) | N,\phi\rangle \langle N, \phi  |$ where $p(N)= \mathrm{tr} \left ( \rho_N \right )$ is the probability for the total number to be $N$. The variance of $N$ is directly obtained from $p(N)$, while the phase dispersion is obtained from
\begin{equation}
\label{GE}
    \langle E \rangle = \sum_{N=0}^\infty p(N) \frac{N}{N+1} \;.
\end{equation}
Let us consider three possibilities for the probability distribution $p(N)$: Poissonian, Gaussian and thermal.

\bigskip
{\it Poissonian statistics.--} 
This is the case of typical number fluctuations of coherent sources where the number variable is distributed as
\begin{equation}
 p(N) = \frac{\bar{N}^N}{N!}e^{-\bar{N}} ,   
\end{equation}
where $\bar{N}$ is the mean number of particles. For this statistics we obtain
\begin{equation}
\Delta^2 N = \bar{N} \;,\qquad  D^2 = 1 - \left( \dfrac{\bar{N}-1+e^{-\bar{N}}}{\bar{N}} \right)^2  \;,
\end{equation}
resulting in no violation of the criteria Eq.~\eqref{NPent} and Eq.~\eqref{NPepr} for all $\bar{N}$.

\bigskip
{\it Gaussian statistics.--} 
This is the case of large enough number of photons so that $N$ can be treated as a continuous variable obeying Gaussian statistics
\begin{equation}
 p(N) = \frac{1}{\sqrt{2 \pi}\Delta N} \exp \left[ - \frac{\left( N - \bar{N} \right)^2 }{ 2 \Delta^2 N} \right] \;,
\end{equation}
where $\bar{N}$ is the mean number of particles, and we assume $\Delta N \ll \bar{N}$. In this limit, to obtain simple expressions, let us consider a series expansion of $N/(N+1)$ in Eq.~\eqref{GE} around $\bar{N}$ to get
\begin{equation}
    \frac{N}{N+1} \simeq  \frac{\bar{N}}{\bar{N}+1} + \frac{N-\bar{N}}{(\bar{N}+1)^2} - \frac{(N-\bar{N})^2}{(\bar{N}+1)^3}  \;, 
\end{equation}
so that after replacing Eq. (\ref{GE}) by an integral we get
\begin{equation}
    \langle E \rangle \simeq \int_{-\infty}^\infty \text{d}N\; p(N) \frac{N}{N+1} \simeq  \frac{\bar{N}}{\bar{N}+1} - \frac{\Delta^2 N}{(\bar{N}+1)^3} \;. 
\end{equation}
Furthermore, since this approximation is valid provided $\bar{N}\gg 1$, we may also consider a series expansion in powers of $1/\bar{N}$ to get to first order 
\begin{equation}
D^2 \simeq \frac{2}{\bar{N}} \;.
\end{equation}
In this case the violation of the criteria Eq.~\eqref{NPent} and Eq.~\eqref{NPepr} depends on the specific values of $\bar{N}$ and $\Delta N$. For $$\bar{N}\gg 1$$ we observe that entanglement is revealed by Eq.~\eqref{NPent} when $\Delta^2 N < \bar{N}/2$, while steering  is revealed by Eq.~\eqref{NPepr} when $\Delta^2 N < \bar{N}/8$. Note here that having $\Delta^2 N < \bar{N}$ corresponds to subPoissonian statistics.

\bigskip
{\it Thermal statistics.--} 
This is the typical case of number fluctuations resulting from thermal light sources, for which
\begin{equation}
 p(N) = \dfrac{1}{\bar{N}+1} \left( \dfrac{\bar{N}}{\bar{N}+1} \right)^N \;,
\end{equation}
where $\bar{N}$ is again the mean number while the variance in this case is 
\begin{equation}
    \Delta^2 N = \bar{N} \left ( \bar{N}+1 \right ).
\end{equation}
The relative-phase dispersion is directly evaluated from
\begin{equation}
    \langle E \rangle = \sum_{N=0}^\infty p(N) \frac{N}{N+1}= 1-\frac{\ln{\left ( \bar{N} + 1 \right )}}{\bar{N}} \;.
\end{equation}
Also in this case there we have that there is no violation of the criteria Eq.~\eqref{NPent} and Eq.~\eqref{NPepr} for any $\bar{N}$.

\section{Conclusions}
In this work we presented criteria to detect entanglement and EPR steering between two bosonic modes, that are based on number and phase measurements. To achieve this, we first presented the operators associated to the latter measurements, together with their uncertainty relation. In particular, to describe fluctuations in the phase we used the concept of dispersion, which is an analogous of the variance for cyclic variables. Then, inspired by the typical EPR sum/difference quadratures, we derived entanglement and steering criteria in terms of total number of particles and relative phase measurements.
These are inequivalent to the typical criteria based on $XP$ quadrature measurements, and therefore they allow to detect quantum correlations in classes of states that are in general different. Moreover, in contrary to criteria requiring to access interference terms such as $\avg{a^\dagger b}$, the criteria we presented can be tested by performing local measurements, allowing to explore the nonlocal character of quantum correltions. We give a few examples where our criteria are violated, showing that they are useful in concrete experimental scenarios. The latter can be easily implemented using optical modes, but also with double-well BECs or mechanical oscillaotrs. Apart from being of fundamental interest, our results could also find application in quantum information protocols, such as number-phase teleportation \cite{Milburn99,Yu00,Cochrane00,Cochrane01,Tran02}.

\section{Acknowledgments}
M.F. acknowledges support from The National Natural Science Foundation of China (Grants No. 11622428 and No. 61675007) and from the Swiss National Science Foundation.
Q. H. thanks the support from the National Natural Science Foundation of China (Grants No. 61675007 and No. 11975026), Beijing Natural Science Foundation (Grant No. Z190005) and the Key R$\&$D Program of Guangdong Province (Grant No. 2018B030329001). L. A. and A. L. acknowledge financial support from Spanish Ministerio de Econom\'ia y Competitividad Project No. FIS2016-75199-P.
L. A. acknowledges financial support from European Social Fund and the Spanish Ministerio de Ciencia Innovaci\'{o}n y Universidades, Contract Grant No. BES-2017-081942.

\end{document}